# Multiple conducting carriers generated in LaAlO$_3$/SrTiO$_3$ heterostructures

S. S. A. Seo,[1*] Z. Marton,[1,2] W. S. Choi,[3] G. W. J. Hassink,[4,5] D. H. A. Blank,[4] H. Y. Hwang,[5,6] T. W. Noh,[3] T. Egami,[1,7] and H. N. Lee[1†]

[1] Materials Science and Technology Division, Oak Ridge National Laboratory, Oak Ridge, TN 37831, USA
[2] Department of Materials Science and Engineering, University of Pennsylvania, Philadelphia, Pennsylvania 19104, USA
[3] ReCOE & FPRD, Department of Physics and Astronomy, Seoul National University, Seoul 151-747, Korea
[4] MESA+ Institute for Nanotechnology, University of Twente, Enschede, NL-7500 AE, The Netherlands
[5] Department of Advanced Materials Science, University of Tokyo, Kashiwa, Chiba 277-8561, Japan
[6] Japan Science and Technology Agency, Kawaguchi, 332-0012, Japan
[7] Department of Physics and Astronomy, University of Tennessee, Knoxville, TN 37996, USA

We have found that there is more than one type of conducting carriers generated in LaAlO$_3$/SrTiO$_3$ heterostructures by comparing the sheet carrier density and mobility from optical transmission spectroscopy with those from *dc*-transport measurements. When multiple types of carriers exist, optical characterization dominantly reflects the contribution from the high-density carriers whereas *dc*-transport measurements may exaggerate the contribution of the high-mobility carriers even though they are present at low-density. Since the low-temperature mobilities determined by *dc*-transport in the LaAlO$_3$/SrTiO$_3$ heterostructures are much higher than those extracted by optical method, we attribute the origin of high-mobility transport to the low-density conducting carriers.

\* seos@ornl.gov

† hnlee@ornl.gov

Interfacial conductance with high-mobility is a scientifically fascinating phenomenon with potential in many technical applications. Recently, observation of metallic transport in a heterointerface between two insulators of LaAlO$_3$ (LAO) and SrTiO$_3$ (STO) has attracted a lot of attention due to its high-mobility (~$10^4$ cm$^2$V$^{-1}$s$^{-1}$) at low temperatures.[1] Following studies have revealed a number of additional features such as interfacial magnetic ordering and electric field controlled superconductivity at the LAO/STO heterointerface.[2,3] At the same time, concerns about the oxygen off-stoichiometry have been raised.[4,5] Thus, characterizing the exact nature of the high-mobility carriers will be an important step toward advancing oxide electronic materials and devices.

In this letter, by optical transmission spectroscopy and *dc*-Hall measurements, we report that there are at least two different kinds of conducting carriers with different mobilities in the LAO/STO heterostructure. Since optical transmission spectroscopy is an *ac*-transport measurement technique, its spectral analysis can provide us with not only qualitative but also quantitative information on the nature of conducting carriers of the LAO/STO system. By comparing the physical quantities extracted from the optical spectroscopic data with the results of conventional *dc*-transport, we suggest that only a fraction of the carriers is responsible for the high-mobility (>$10^3$ cm$^2$V$^{-1}$s$^{-1}$) observed in the LAO/STO heterostructure while the majority of conducting carriers have low-mobility (~10 cm$^2$V$^{-1}$s$^{-1}$).

We grew 125 unit-cell-thick LAO thin films on TiO$_2$-terminated STO (001) substrates by pulsed laser deposition (PLD).[6] Figure 1(a) shows the metallic sheet resistance, $R_{xx}$, of an as-grown sample as a function of temperature. Another sample, which was prepared under the identical condition followed by *in-situ* post-annealing in a higher oxygen partial pressure (P$_{O2}$ = 50 mTorr) for 10 min at the growth temperature, shows a highly resistive and insulating behavior, which is consistent with the previous reports.[4,5] The difference in optical transmission of the LAO/STO samples is also distinguishable by eye: The metallic as-grown sample has a dark bluish color while the post-annealed sample is as transparent as an insulating STO substrate (see inset of Fig. 1(a)).

Figure 1(b) shows optical transmittance spectra ($T(\omega)$) of the LAO/STO samples. The as-grown (metallic) sample has overall a low transmittance in the photon energy region of 0.5 - 3.3 eV. On the other hand, the post-annealed (insulating) sample has a high transmittance reaching 80 % below 3.2 eV. The common steep drop in transmittance at 3.2 eV originates from the $O_{2p} \rightarrow Ti_{3d}$ charge-transfer transition of the STO substrates. The low transmittance from the as-grown sample at lower photon energies is a signature of conducting carriers, which will be discussed below. Interestingly, the post-annealed sample shows even higher transmittance than a bare STO substrate in the measured photon energy region due to the anti-reflection effect.[7] To identify the detailed spectral features of the as-grown sample, a superlattice of $[(LaTiO_3)_1/(LAO)_5]_{20}$ (a total of 120 u.c. in thickness) on a STO (001) substrate ([LTO/LAO]/STO) was also tested. Note that both as-grown LAO/STO and [LTO/LAO]/STO show a very similar spectral shape (Fig. 1(b)): Transmittance decreasing to zero at the lowest photon energy (~0.5 eV) and three weak absorption bands at 1.7, 2.4, and 2.9 eV (marked with dashed vertical lines) are observed for both samples. However, after post-annealing under oxygenating conditions (flowing $O_2$, at 400 °C for 1 hr), the [LTO/LAO]/STO sample also becomes transparent and shows quite flat $T(\omega)$ (Fig. 1(b)). Interestingly, however, x-ray diffraction does not reveal any structural changes (data not shown here), suggesting that the spectral features of as-grown samples come mainly from the STO substrate. To clarify this point, we grew a STO film (250 u.c.) on STO (001) in the same conditions as those for LAO/STO. By comparing its $T(\omega)$ with other heterostructures, it is clear that the above-mentioned spectral features originate not from LAO but from STO (Fig. 1(b)). Below we focus our discussion only on features of the absorption spectra that are directly correlated with the transport properties of conducting carriers, rather than the origins of the absorption bands[6] at the finite photon energies.

Figure 2 shows simulated $T(\omega)$ for an LAO/STO sample.[6] Increasing the sheet carrier density ($n_s$) from $10^{14}$ to $10^{19}$ cm$^{-2}$ while fixing the mobility ($\mu$) at 4 cm$^2$V$^{-1}$s$^{-1}$ results in systematically-decreased $T(\omega)$ and the variation becomes larger particularly at lower photon energies, as shown in Fig. 2(a). By comparing these simulated spectra with the experimental spectrum of LAO/STO, we can estimate that $n_s$ is about $3\times10^{17}$ cm$^{-2}$. In addition, the evolution of $T(\omega)$ as a function of $\mu$ at a fixed $n_s = 3\times10^{17}$ cm$^{-2}$ is shown in Fig. 2(b). At higher mobilities ($\mu = 10-1000$ cm$^2$V$^{-1}$s$^{-1}$), there is a drastic change in the curvature of the spectra at lower photon energies while the spectra with lower mobilities ($\mu = 0.01-1$ cm$^2$V$^{-1}$s$^{-1}$) exhibit a rather moderate change at the same energies. Thus, one can estimate $n_s$ and $\mu$ of the conducting carriers generated with reasonable confidence. Interestingly, the simulated spectra at $\mu = 10000$ and $0.01$ cm$^2$V$^{-1}$s$^{-1}$ are very similar to each other for different reasons: The former has a very narrow Drude absorption feature with a small scattering rate, which lies below our photon energy range, and the latter has a rather broad featureless shape of the Drude absorption over a wide region of photon energies.

Since the high $\mu$ values of free carriers in LAO/STO at low temperature are of interest, we measured temperature-dependent $T(\omega)$, as shown in Fig. 3(a). Note that the most pronounced change in transmittance is observed at lower photon energies (approximately 0.5−1.5 eV) while the weak absorption bands centered at 1.7 and 2.4 eV show subtle blue-shifts.

Figure 3(b) summarizes $n_s$ and $\mu$ extracted from the model fits in Fig. 3(a). For comparison, we also plot the data obtained by *dc*-Hall measurements. As shown in the inset, the values of $n_s$ measured by both optical and *dc* techniques are in good agreement and remain almost unchanged ($1-3\times10^{17}$ cm$^{-2}$) with temperature. Although $n_s \approx 10^{17}$ cm$^{-2}$ are consistently observed in LAO/STO heterostructures,[1,4,5] it gives a very high carrier density if one assumes that the carriers are confined within a few hundreds nanometers or less.[8] Such a high carrier density has been attributed to the fact that the growth of oxide thin films under reducing conditions by highly energetic PLD can effectively generate a high concentration of electron donors, *e.g.* oxygen vacancies, in underlying STO as reported recently by Herranz *et al*.[9]

Note that the mobility at low temperature, $\mu$(10 K), determined by spectral analyses is enhanced by about a factor of two as compared to the room-temperature value, while the *dc*-transport $\mu$ is increased enormously by about three orders of magnitude, as shown in Fig. 3(b). In inhomogeneous materials, the *dc*-transport may show lower conductivity than the optically measured one since the *dc*-transport measurement strongly depends on the connectivity of the conducting channels while the optical spectroscopic measurement does not.[10] However, what we have observed here strongly suggests that the carriers that contribute dominantly to the *dc*-transport and optical spectroscopy are not the same, and there exist at least two kinds of carriers.[6] One that contributes mostly to the *dc*-transport has high-mobility, low-density, and strong temperature dependence. The other one that is seen by optical spectroscopy has low-mobility, high-density, and weak temperature dependence. If we assume that there exist multiple, parallel conducting channels within STO, the sheet conductivity can be expressed as $S_{xx}(\equiv 1/R_{xx}) = n_{s1}e\mu_1 + n_{s2}e\mu_2 + \cdots$, then both results from *dc*-transport and optical spectroscopy can be understood consistently. Thus *dc*-conductivity is dominated by high-mobility carriers. On the other hand, optical spectra are dominated by high-density carriers, regardless of their mobility.

At this moment, it is still unclear which part of STO, *i.e.* the region close to or far from the interface, has higher $\mu$ carriers, since both optical spectroscopy and *dc*-transport are macroscopic bulk-sensitive techniques. Nevertheless, the origin of low-density carriers with high $\mu$ at low temperature seems to be related to the quantum paraelectric behavior of STO. It is well known that the dielectric permittivity of STO diverges as it approaches the incipient-ferroelectric transition at low temperature.[11] The increased dielectric permittivity can effectively screen the scattering sources such as defects from mobile electron carriers and thus their $\mu$ can be enormously increased. Recently, Copie *et al.*[12] reported that carriers can be confined in a very narrow region near interface even at low temperature by considering reduction of the dielectric permittivity of STO under a strong electric field. Hence, a reasonable picture might be that high-density low-mobility carriers are confined near the interface region while low-density high-mobility carriers can be distributed far from the interface. Since the low-density high-mobility carriers are distributed in a high dielectric permittivity region, the electric field control of the conductivity[2] might have been realized effectively in this type of heterostructures. To further clarify the spatial distribution and $\mu$ profile of the conducting carriers, a microscopic transport technique such as quadra-probe scanning tunneling microscopy at low temperature[13] on the cross-sectional heterointerface might be a useful approach. Regarding the origin of the multiple carriers, we also need further experimental investigations even though the finding reported here – the conduction in this heterostructure can be intrinsically associated with more than one type of carriers – has been previously proposed.[3,14]

In summary, transport properties of conducting carriers generated at the polar/non-polar interface in LAO/STO heterostructures have been investigated by optical transmission spectroscopy and *dc*-transport. We demonstrate that $n_s \approx 10^{17}$ cm$^{-2}$, which is consistent with the result of *dc*-transport measurements, can be estimated by analyzing the temperature-dependent optical transmittance spectra. However, while the *dc*-transport mobility shows a huge increase by three orders of magnitude as temperature decreases, the optically-probed mobility only approximately doubles. The discrepancy can be understood by the assumption that low-density high-mobility carriers are generated in the metallic LAO/STO heterostructures in addition to high-density low-mobility carriers.

We thank I. Ivanov and H. M. Christen for useful discussions. This work was sponsored by the Division of Materials Sciences and Engineering, U. S. Department of Energy, a nanotechnology program of the Dutch Ministry of Economic Affairs (NanoNed), and National Science Foundation by DMR04-04781.

# References

[1] A. Ohtomo and H. Y. Hwang, Nature **427**, 423 (2004).

[2] S. Thiel, G. Hammerl, A. Schmehl, C. W. Schneider, and J. Mannhart, Science **313**, 1942 (2006).

[3] A. Brinkman, M. Huijben, M. van Zalk, J. Huijben, U. Zeitler, J. C. Maan, W. G. van der Wiel, G. Rijnders, D. H. A. Blank, and H. Hilgenkamp, Nat. Mater. **6**, 493 (2007).

[4] A. Kalabukhov, R. Gunnarsson, J. Borjesson, E. Olsson, T. Claeson, and D. Winkler, Phys. Rev. B **75**, 121404 (2007).

[5] G. Herranz, M. Basletic, M. Bibes, C. Carretero, E. Tafra, E. Jacquet, K. Bouzehouane, C. Deranlot, A. Hamzic, J.-M. Broto, A. Barthelemy, and A. Fert, Phys. Rev. Lett. **98**, 216803 (2007).

[6] For details, see the Supplemental Material.

[7] The refractive index of LAO is smaller than that of STO. For the anti-reflection effect, see G. R. Fowles, *Introduction to Modern Optics*, $2_{nd}$ ed. (Dover Publication Inc., New York, 1989).

[8] If the effective thickness of the carriers is about 100 nm, then the carrier density becomes about $10^{22}$ cm$^{-3}$. One electron carrier per Ti ion gives $1.68\times10^{22}$ cm$^{-3}$ in carrier density. Also note that half-electron per interfacial unit-cell ($n_s \approx 3.3\times10^{14}$ cm$^{-2}$) is required to avoid the polar catastrophe.

[9] G. Herranz, M. Basletic, O. Copie, M. Bibes, A. N. Khodan, C. Carretero, E. Tafra, E. Jacquet, K. Bouzehouane, A. Hamzic, and A. Barthelemy, Appl. Phys. Lett. **94**, 012113 (2009).

[10] K. H. Kim, J. Y. Gu, H. S. Choi, D. J. Eom, J. H. Jung, and T. W. Noh, Phys. Rev. B **55**, 4023 (1997).

[11] T. Sakudo and H. Unoki, Phys. Rev. Lett. **26**, 851 (1971).

[12] O. Copie, V. Garcia, C. Bodefeld, C. Carretero, M. Bibes, G. Herranz, E. Jacquet, J.-L. Maurice, B. Vinter, S. Fusil, K. Bouzehouane, H. Jaffres, and A. Barthelemy, Phys. Rev. Lett. **102**, 216804 (2009).

[13] T.-H. Kim, Z. Wang, J. F. Wendelken, H. H. Weitering, W. Li, and A.-P. Li, Rev. Sci. Instrum. **78**, 123701 (2007).

[14] Z. S. Popovic, S. Satpathy, and R. M. Martin, Phys. Rev. Lett. **101**, 256801 (2008).

[15] Regarding *m*\* of STO-based conducting materials, see M. Ahrens, R. Merkle, B. Rahmati, and J. Maier, Physica B: Cond. Mat. **393**, 239 (2007) and S. S. A. Seo, W. S. Choi, H. N. Lee, L. Yu, K. W. Kim, C. Bernhard, and T. W. Noh, Phys. Rev. Lett. **99**, 266801 (2007).

**Figure captions**

FIG. 1. (Color online) (a) Sheet resistance of LAO (125 u.c.) films on STO substrates. Inset: photographs of as-grown and post-annealed LAO/STO samples on 5×5 mm$^2$ mesh paper. (b) Room temperature $T(\omega)$ of LAO films on STO, [LTO/LAO] superlattices on STO, a STO film on STO, and a bare STO substrate.

FIG. 2. (Color online) Simulated $T(\omega)$ of a LAO (50 nm) thin film on STO substrate. (The experimental data of the LAO/STO sample is also shown for comparison.) (a) Simulated spectra with $n_s$ from $1\times10^{14}$ to $1\times10^{19}$ cm$^{-2}$ at a fixed mobility $\mu = 4$ cm$^2$V$^{-1}$s$^{-1}$ and an effective mass $m^* = 1.8m_e$.[15] Green lines are calculated curves for $n_s$ from $2\times10^{17}$ to $9\times10^{17}$ cm$^{-2}$. (b) Simulated spectra with various mobility values $\mu = 10^{-2}$–$10^4$ cm$^2$V$^{-1}$s$^{-1}$ at $n_s = 3\times10^{17}$ cm$^{-2}$. Transmittance measurement at very low photon energy (shaded region) is practically hindered by the Restrahlen band of infrared-active phonons of the STO substrate.

FIG. 3. (Color online) (a) Temperature-dependent $T(\omega)$ of LAO/STO and their model-fit curves at each temperature. The arrows point the weak absorption bands showing a blue-shift. (b) The mobility as a function of temperature extracted from *dc*-resistivity and Hall measurements as well as optical transmission spectroscopy. Inset shows $n_s$ as a function of temperature.

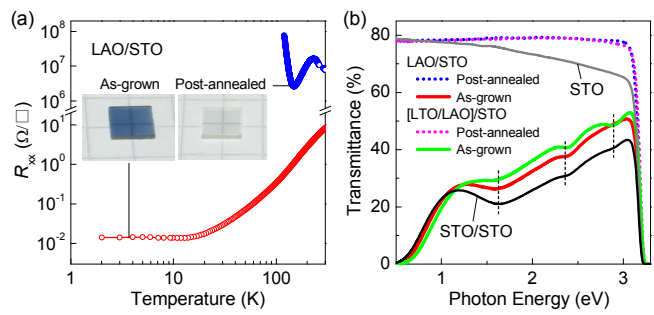

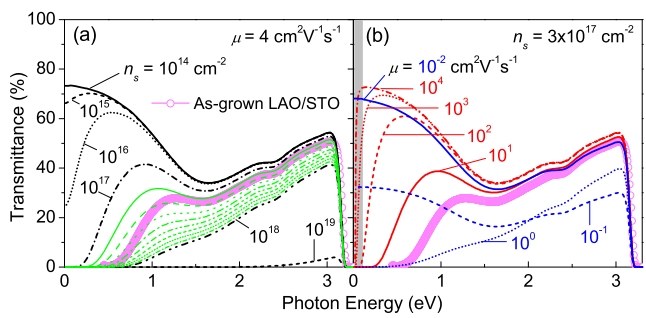

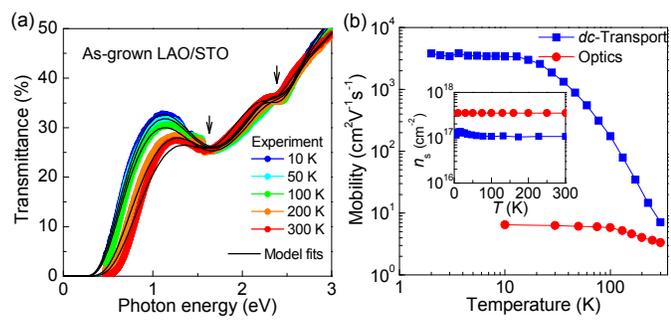

*Supplemental Material*

## Multiple conducting carriers generated in LaAlO$_3$/SrTiO$_3$ heterostructures

S. S. A. Seo *et al.*

### 1. Sample growth conditions

125 unit-cell thick LaAlO$_3$ (LAO) thin films (approx. 50 nm in thickness) were grown on TiO$_2$-terminated SrTiO$_3$ (STO) (001) substrates by pulsed laser deposition with *in-situ* monitoring of the specular spot intensity oscillations of reflection high-energy electron diffraction. While LAO has a wide window of epitaxial growth, we have chosen growth conditions which effectively produce highly conducting carriers in the LAO/STO system (P$_{O2}$: 2×10$^{-6}$ Torr, temperature: 700 °C, and KrF ($\lambda$ = 248 nm) excimer-laser fluence on a target: 1.2 Jcm$^{-2}$). A superlattice sample of [(LaTiO$_3$)$_1$/(LAO)$_5$]$_{20}$ on a STO (001) substrate was grown in P$_{O2}$: 1×10$^{-6}$ Torr and 700 °C.

### 2. Measurement details

Optical transmittance spectra ($T(\omega)$) were measured using a Cary5000 (Varian Co.) spectrophotometer in a normal incidence geometry at room temperature. To avoid blocking/scattering light beams, the backside of substrates was evenly polished by fine (0.25 micron) diamond-powder pastes until being optically-flat while the front surface (LAO film side) was protected by crystal glue. Low-temperature $T(\omega)$ was measured using a Cary5G (Varian Co.) spectrophotometer with a home-built optical cryostat mount equipped with an Oxford Optistat with quartz windows. Temperature-dependent *dc*-transport measurements were done using a Quantum Design Physical Property Measurement System.

## 3. Remarks on the absorption bands at finite photon energies

The weak absorption bands in $T(\omega)$ of the as-grown samples at finite photon energies (1.7, 2.4, and 2.9 eV) are responsible for the dark bluish color of these samples. It is unlikely that such wavy spectra come from the optical interference between the thin film surface and interface because the refractive indices of the LAO film ($N_{LAO}$) and the STO film ($N_{STO}$) are ~2.0 and ~2.3, respectively, in this photon energy region, and $N_{LAO} \cdot 50$ (nm) $\neq N_{STO} \cdot 100$ (nm). The three weak absorption bands are located at exactly the same energies as absorptions in reduced STO bulk crystals (1.7, 2.4, and 2.9 eV), again suggesting that their origin is STO. The absorption peak at 1.7 eV is known to be enhanced by reducing a STO crystal further[1] while the one at 2.4 eV is independent of the free carrier concentration and might originate from the excitation of an electron trapped by an oxygen vacancy, *i.e.*, in an $F_1$ center.[2] Finally, the 2.9 eV absorption has been observed in reduced, Fe-doped STO crystals, even though it has not been specifically attributed to a certain defect,[1,3,4] and could be eliminated by Ar/H$_2$ annealing.[5]

## 4. Optical model used for extracting the sheet carrier density and mobility

We used a model of an isotropic film on an isotropic substrate[6] to simulate $T(\omega)$ for LAO/STO samples based on the intensity transfer matrix method (ITMM).[7] We used the dielectric functions of the LAO thin film, which were independently measured by optical ellipsometry, and the dielectric function of the interfacial metallic region ($\widetilde{\varepsilon}_{IF}(\omega)$) within the effective thickness ($d^*$) by introducing optical parameters:

$$\widetilde{\varepsilon}_{IF}(\omega) = \widetilde{\varepsilon}^{STO}(\omega) - \frac{\omega_p^2}{\omega^2 + i\omega\gamma} + \sum_j \frac{S_j \omega_{0j}^2}{\omega_{0j}^2 - \omega^2 - i\omega\Gamma_j}, \qquad (1)$$

where $\widetilde{\varepsilon}^{STO}(\omega) [\equiv \varepsilon_1^{STO}(\omega) + i\varepsilon_2^{STO}(\omega)]$ is the complex dielectric function of STO, which was obtained from an STO bare substrate, and $\omega_p$ and $\gamma$ are the plasma frequency and the scattering rate in their simple Drude form, which are correlated with the sheet carrier density ($n_s = \omega_p^2 m^* d^* / 4\pi e^2$, where $m^*$ is the effective mass[8] and $e$ is the electron charge) and the mobility ($\mu = \frac{e}{m^*\gamma}$) of free carriers, respectively. To take into account the effect from the optical absorptions at finite photon energies, a third term containing Lorentz oscillators is introduced in our model fits and its parameters are listed in Table 1.

Since our samples are in a weakly absorbing regime ($T(\omega) \neq 0$), $T(\omega)$ can be approximated as:

$$T(\omega) \equiv \frac{I(\omega)}{I_0(\omega)} \propto \exp(-\alpha(\omega) \cdot d^*) \approx \exp\left(-\frac{4\pi e^2}{c\sqrt{\varepsilon_1^{STO}(\omega)}} \cdot \frac{n_s}{m^*} \cdot \frac{\gamma}{\omega^2 + \gamma^2}\right), \qquad (2)$$

where $\alpha(\omega)$ is the absorption coefficient of the metallic region and $c$ is the light velocity. Note that $T(\omega)$ is independent of $d^*$. Hence, we can determine $n_s$ and $\mu$ without assuming the depth-profile of the metallic carriers. In our simulation, we assume a flat top-surface, an abrupt interface between LAO and STO, and a homogeneous layer of $\widetilde{\varepsilon}_{IF}(\omega)$ without it forming an additional well-defined interface within STO. This is a good approximation for $T(\omega)$ simulation since the conducting electrons in a metallic LAO/STO sample are gradually distributed within the STO substrate, as recently reported.[9] When we examine another model by inserting a few micrometer-thick layers of $\widetilde{\varepsilon}_{IF}(\omega)$ forming an abrupt interface within STO, the calculated $T(\omega)$ shows interference patterns from these layers, which is not consistent with our experimental data.

**Table 1.** Lorentz oscillator parameters for the weak absorptions at 1.7, 2.4, and 2.9 eV used for the simulation in Fig. 2.

| $J$ | $\omega_0$ (eV) | $S$ | $\Gamma$ (eV) |
|---|---|---|---|
| 1 | 1.7 | $1.03\times10^{-4}$ | 1.05 |
| 2 | 2.4 | $1.10\times10^{-5}$ | 0.60 |
| 3 | 2.9 | $9.00\times10^{-6}$ | 1.00 |

## 5. An estimation of low-density carriers

First, we show how a single type carrier model cannot be compatible with our data. If we substitute the single values of $n_s$ and $\mu$ determined from the *dc*-transport measurement into a spectral simulation, then the simulated transmittance at 10 K is much higher in the lower photon energy region than the experimentally measured one, as shown in Fig. S1 (F1).

If we assume two-type carriers (high-density low-mobility carriers and low-density high-mobility carriers), then we can find a good fit on an experimental spectrum. Figure S1 (F2) shows a simulated transmittance spectrum at 10 K, in which parameters have carriers of $n_{s1} = 3.53\times10^{17}$ cm$^{-2}$ and $\mu_1 = 6.5$ cm$^2$V$^{-1}$s$^{-1}$, and additional carriers of $n_{s2} = 1.30\times10^{16}$ cm$^{-2}$ and $\mu_2 = 34,000$ cm$^2$V$^{-1}$s$^{-1}$. The combination of $n_{s2}$ and $\mu_2$ is not a unique solution but one of several possible parameters that can provide us with a good fit since optical spectrum is dominated by $n_{s1}$ and $\mu_1$. Note that $n_{s2} \approx 10^{16}$ cm$^{-2}$ is just around at the upper limit and may have smaller values.

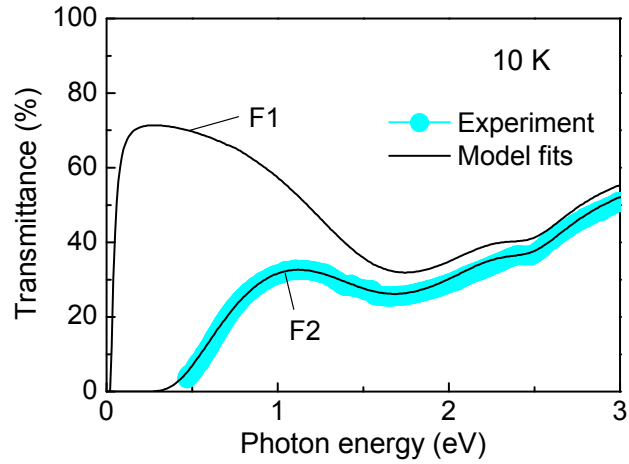

**FIG. S1** (Color online) Simulated transmission spectra based on two parameter-sets for the 10 K data. For **F1**, the values ($n_s = 1.30\times10^{17}$ cm$^{-2}$, $\mu = 3,400$ cm$^2$V$^{-1}$s$^{-1}$) obtained by *dc*-transport were used, while high-density low mobility carriers ($n_{s1} = 3.53\times10^{17}$ cm$^{-2}$, $\mu_1 = 6.5$ cm$^2$V$^{-1}$s$^{-1}$) and low-density high mobility carriers ($n_{s2} = 1.30\times10^{16}$ cm$^{-2}$, $\mu_2 = 34,000$ cm$^2$V$^{-1}$s$^{-1}$) were used for **F2**. An experimental spectrum of LAO/STO is shown for comparison.

# References

[1] C. Lee, J. Destry, and J. L. Brebner, Phys. Rev. B **11**, 2299 (1975).

[2] W. S. Baer, Phys. Rev. **144**, 734 (1966).

[3] R. L. Wild, E. M. Rockar, and J. C. Smith, Phys. Rev. B **8**, 3828 (1973).

[4] K. W. Blazey and H. Weibel, J. Phys. Chem. Solids **45**, 917 (1984).

[5] B. Jalan, R. Engel-Herbert, T. E. Mates, and S. Stemmer, Appl. Phys. Lett. **93**, 052907 (2008).

[6] R. M. A. Azzam and N. M. Bashara, *Ellipsometry and Polarized Light*. (North-Holland, Amsterdam, 1987).

[7] T. W. Noh, P. H. Song, S.-I. Lee, D. C. Harris, J. R. Gaines, and J. C. Garland, Phys. Rev. B **46**, 4212 (1992).

[8] In our simulation, we assumed $m^* = 1.8 m_e$ for STO.

[9] M. Basletic, J. L. Maurice, C. Carretero, G. Herranz, O. Copie, M. Bibes, E. Jacquet, K. Bouzehouane, S. Fusil, and A. Barthelemy, Nat. Mater. **7**, 621 (2008).